\documentclass[%
aip,
sd,%
amsmath,amssymb,
reprint,%
]{revtex4-1}
\usepackage[colorlinks=true,
linkcolor=red,
urlcolor=black,
citecolor=blue]{hyperref}
\usepackage{graphicx}
\usepackage{epstopdf}
\usepackage{dcolumn}
\usepackage{bm}

\begin{document}
\preprint{AIP/123-QED}
\title{Coexisting synchronous and asynchronous states in locally coupled array of oscillators by partial self-feedback control}
	\author{Bidesh K. Bera}
	\author{Dibakar Ghosh}
	\affiliation{Physics and Applied Mathematics Unit, Indian Statistical Institute, Kolkata-700108, India}
	\email{dibakar@isical.ac.in}
	\author{Punit Parmananda}
		\affiliation{Department of Physics, Indian Institute of Technology Bombay, Powai, Mumbai-400 076, India}
	\author{G. V. Osipov}
		\affiliation{Department of Control Theory, Nizhni Novgorod State University, Gagarin Avenue 23, 606950, Nizhni Novgorod, Russia}
	\author{Syamal K. Dana}
	\affiliation{Department of Mathematics, Jadavpur University, Kolkata 700032, India}


	\date{\today}
	
	\begin{abstract}
		We report the emergence of coexisting synchronous and asynchronous subpopulations of oscillators  in one dimensional arrays of identical oscillators by applying a self-feedback control. When  a self-feedback is applied to  a  subpopulation of the array,   similar to chimera states, it splits into two/more sub-subpopulations coexisting in coherent and incoherent states for a range of self-feedback strength.  By tuning  the coupling  between the  nearest neighbors and the amount of self-feedback in the perturbed subpopulation,  the size of the coherent and the incoherent  sub-subpopulations in the array can be controlled, although the exact  size  of them is unpredictable. We present numerical evidence using the Landau-Stuart (LS) system and the Kuramoto-Sakaguchi (KS) phase model.
		
	\end{abstract}
	
	\pacs{05.45.Xt, 05.45.Pq}
	\maketitle
	\begin{quotation}
		{\bf  The symmetry breaking of a coherent population of identical oscillators under non-local, local and global coupling, into coexisting coherent and incoherent subpopulations is a strange behavior that has been discussed in recent literature of collective behavior of many dynamical systems. Such a self-organized behavior of a homogeneous population of oscillators is called chimera states. We observe emergence of a similar type of splitting  into  subsets of coexisting synchronous and asynchronous oscillators, when we apply  a self-feedback to a coherent subset of oscillators in a locally coupled  open chain of identical oscillators. Interestingly, the unperturbed subpopulation remains coherent for higher values of feedback strength. Most encouragingly, it shows a direction how the size of the coherent and the incoherent subpopulations in the chimera pattern can be monitored by playing with the feedback strength. We provide numerical evidence in two such arrays using the Landau-Stuart (LS) system and the Kuramoto-Sakaguchi (KS) phase model. }
	\end{quotation}

\section{Introduction}
Besides synchronization and clustering in a network of oscillators,  the emergence of chimera states in identical oscillators under nonlocal coupling \cite{kuramoto, strogaz_prl, chi_rev} and, even for global coupling \cite{global} and nearest neighbor coupling \cite{bidesh1,local,bidesh2, bidesh3} have received significant attention  in recent years. The chimera states emerge as a surprising symmetry breaking in a homogeneous population of identical oscillators into two coexisting synchronous and asynchronous subpopulations; the mechanism  of such a symmetry breaking is yet to be unraveled completely.  However, recent studies \cite{natcom_laser,glob_laser, minimal_glob, powerlaw_eco} provided some clues on the possible mechanisms of such a symmetry breaking. Chimera states were first observed by Kuramoto and Battogtokh \cite{kuramoto} in a network of nonlocally coupled identical phase oscillators. 
In the real world and man-made systems,  power grid network \cite{power1,power2} and social network, \cite{juan} chimera or chimera-like states have been observed. The signatures of  the chimera-like behavior were noticed  in mammals and birds when they are engaged in unihemispheric slow-wave sleep \cite{ratten1,ratten2}.  
It has been mimicked in laboratory experiments \cite{nature1,nature2,pnas,kapitaniak} too. Some recent reports  have also identified chimera states in neuronal systems \cite{bidesh1, bidesh2, sci_rep, rev1}  and superconducting Josephson junctions \cite{lazarides,mishra}, which encourage  a search for future application possibilities. Recent studies have also reported emergence of chimera states in multiplex networks \cite{multiplex}. The effect of initial conditions on chimera states recently investigated using  a basin stability approach \cite{bs_chimera}. We raise one important question here: how to  control chimera pattern \cite{Frasca, Erik, Sieber, hovel16}, mainly, how to restrict the pattern to a targeted population or to a selected set of oscillators? The answers definitely will make the chimera states functionally and practically more relevant \cite{Erik}. Since chimera states are an emergent property of a large population, controlling  the size or position of the coherent or incoherent subpopulations in a large network of oscillators is not an easy task. 
\par We adopt a simpler approach, atypical of the conventional approaches \cite{kuramoto, strogaz_prl, chi_rev, global, bidesh1,local,bidesh2, bidesh3,epl_pers}, 
but the traditional self-feedback technique to create and control chimera-like coexisting coherent and incoherent subpopulations in an open chain of identical oscillators. The size of the coherent and the incoherent  subpopulations  is  an emergent property and usually undetermined, however, we find that it is controllable by the feedback control mechanism.   We explain, in this paper, how chimera-like states emerge in an open chain of oscillators by the application of partial self-feedback to a population of identical oscillators; a subpopulation is perturbed by self-feedback and thereby the chimera pattern  become controllable to an extent.
 
The self-feedback mechanism is a well-known engineering technique that has been used, in the past, for controlling complexity \cite{yorke, pla,prl} in dynamical systems including brain dynamics \cite{tass}. 
Possibly, the strategy of controlling chimera states is known to animals who showed signatures of chimera states in a unihemispheric sleeping phase. It is known that mammals could keep  sleeping with one eye open and the other closed and can switch the open eye when necessary depending upon the location of an intruder and accordingly, switch on or off the two hemispheres of the brain and thereby exchange the positions of the coherent and incoherent subpopulations of neurons in the brain. We apprehend that a possible internal feedback mechanism works in favor of such a control mechanism and this leads us to adopt the feedback control technique to create chimera-like states in an open chain of oscillators and to search for a possible control technique.  Parmananda et al. \cite{punit} showed earlier that application of linear-feedback signal can control synchrony in an ensemble of oscillators. This study leads us to  attempt the self-feedback approach and in the process, we address one important question: what are the locations and the sizes of the coherent and the incoherent subpopulations in a chimera pattern, and  are they be controllable?  This answers another question: how to stop a spreading of the incoherent subpopulation, i.e.,  to keep under control the subpopulation of dynamical units moving out of the coherent state to the incoherent state?  
\par We consider all identical oscillators in the open chain and assume them interacting  with nearby neighbors by linear diffusive coupling. The chain emerges into complete coherence for a choice of coupling strength above a threshold. Then we perturb a subpopulation of the dynamical units by adding an equal amount of self-feedback to each of them, and thereby maintain homogeneity of the subpopulation only. 
 Interestingly, the homogeneous subpopulation splits into  coexisting coherent and incoherent sub-subpopulations for an appropriate amount of self-feedback. This symmetry breaking although limited to the perturbed subpopulation only, is analogous to the typical chimera states. 
 The fraction of coherent and incoherent oscillators within the subpopulation is an emergent property of the system. The sizes of the sub-subpopulations  are undetermined,  however, found   controllable by playing with the feedback strength.  We can increase or decrease the size of the coherent and incoherent subpopulations, by the feedback control, although we cannot target a desired number of  the subpopulations. Our numerical study reveals the nature of escape of the individual oscillators from the coherent state to the incoherent state or vice versa by the feedback control. 
 We use the LS system and the KS phase model as individual nodes in  open chain networks. Based on the position of  the application of self-feedback to oscillators, we investigate three different cases, namely, one symmetric feedback scheme and two asymmetric  schemes. 

\section{Network of Landau-Stuart oscillators}
\par We consider $N$ identical LS oscillators coupled via nearest neighbor diffusive coupling to form a one dimensional open array. We first assure a coherent state in the network by applying an appropriate strength of the local coupling and then apply self-feedback to a selected (number and position) subpopulation of dynamical units. The proposed network is 
\begin{equation}
\begin{array}{lcl}
\dot x_i=(1-p_i)x_i-\omega y_i+\epsilon (x_{i+1}+x_{i-1}-2x_i)+f(i)kx_i\\\\
\dot y_i=(1-p_i)y_i+\omega x_i, \;\;\;\;\;\; i=1,2,...,N
\end{array}
\end{equation}
where $p_i=x_i^2+y_i^2$, $\omega $ is the natural frequency of the individual oscillator, $\epsilon$ is the local coupling strength between the neighbors and $k$ is the feedback strength;
$\omega=2.0$ for all the considered cases. $f(i)=1$ for the $i^{th}$ oscillator when a self-feedback is applied  and $f(i)=0$ for no feedback to it. For one-dimensional open arrays, $x_0=x_1$ and $x_{N+1}=x_N$. In absence of any self-feedback (i.e., $k=0$), all the dynamical units are in completely coherent state for a critical value of $\epsilon$. 
Once  each unit of the coherent subpopulation is perturbed by an equal amount of self-feedback with an appropriate strength, the perturbed subpopulation splits into two coexisting coherent and incoherent sub-subpopulations. We emphasize the position of the dynamical units where the self-feedback is to be induced in the subpopulation of the array. We explore three different  cases, depending on the position of the oscillators where the self-feedback is  applied, (1) to the middle of an array with no feedback to identical number of  units on two sides, (2) to  one side of the population, equally divided, (3) to unequal number of units without feedback on two sides.
\par We numerically integrate the proposed network using  the $5^{th}$-order Runge-Kutta-Fehlberg algorithm with a time step size 0.01. We use the following initial condition: $x_m=0.001 (N/2-m)$, $x_n=0.0015 (n-N/2)$, $y_m=0.002 (N/2-m)$, $y_n=0.0012 (n-N/2)$  (where $m=1\cdots100$ and $n=101\cdots200$, $N=m+n$) with added small random fluctuations. We explore the spatiotemporal dynamics of the network  for the above three cases with varying  local coupling strength $\epsilon$ and self-feedback strength $k$.  We identify the parameter space in the $\epsilon$-$k$ plane where  chimera-like states emerge.  

\subsection{Symmetric Feedback: Central population}
We consider the first case  (1) where the self-feedback of identical strength is applied to all the oscillators except the first and the last  units in the chain. The feedback scheme is defined by $f(i)=1$ for $i=2,3,...,N-1$ and $f(1)=f(N)=0$.
\begin{figure}[ht]
	\centering
	\includegraphics[width=0.45\textwidth]{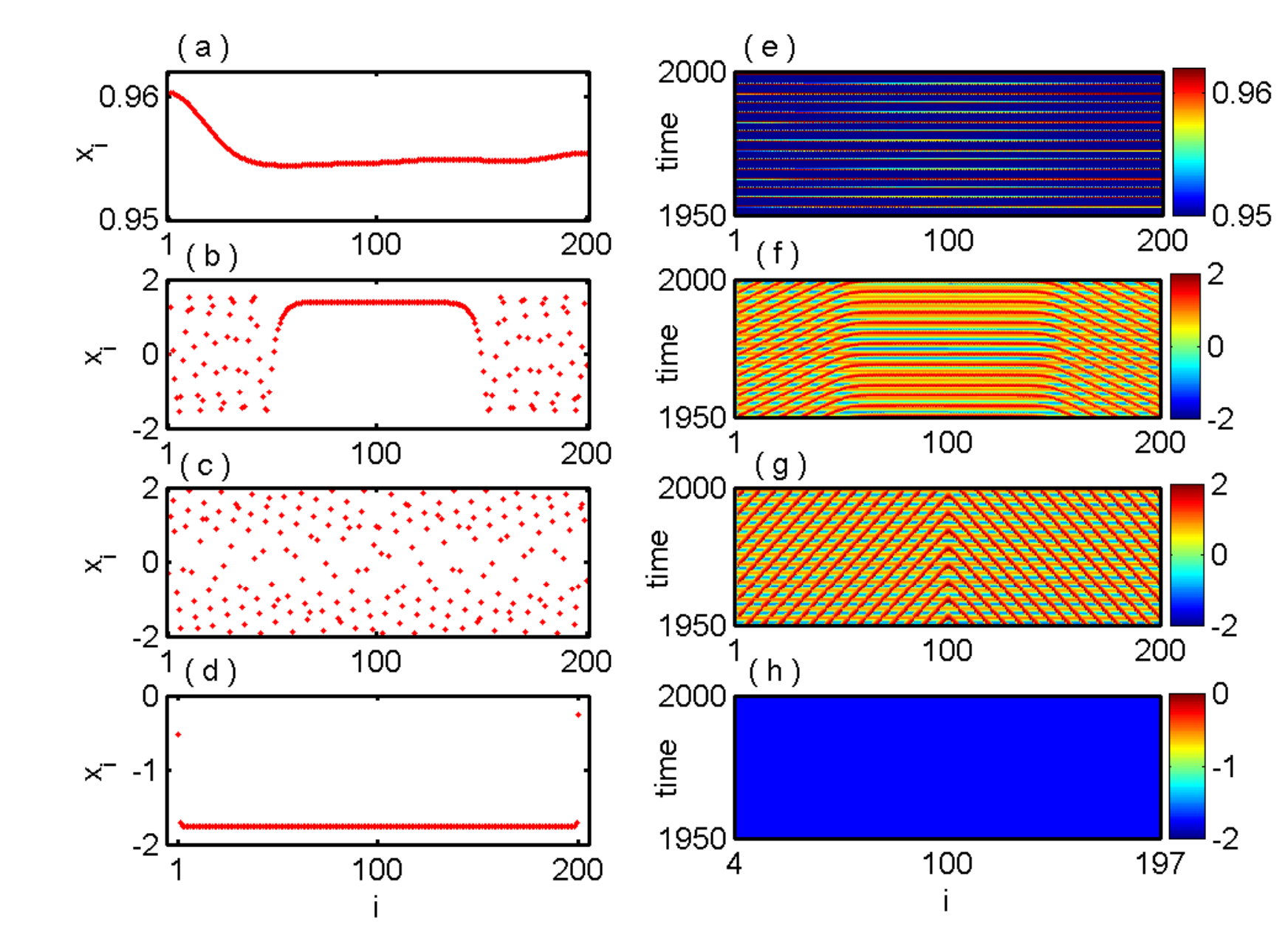} \\[-0.2cm]
	\caption{LS System: snapshots of $x_i$ (left panel) at $t=1980$ for a coupling strength $\epsilon=0.15$, $N=200$ and (a) $k =0$, coherent state, (b) $k =2.0$,  chimera-like state, (c) $k=3.5$, incoherent state, and (d) $k =4.5$, homogeneous steady state.  Spatiotemporal evolutions (right panel) of all the oscillators are shown in (e)-(i) that correspond to figures (a)-(d) respectively. $f(1)=f(N)=0$ and $f(2)=f(3)=...=f(N-1)=1.0$.}
	\label{bothsym}
\end{figure}

Without any self-feedback ($k=0$), we realize a coherent state for a coupling strength, $\epsilon=0.15$ and then apply the self-feedback ($k>$0).  Figure ~\ref{bothsym} shows  snapshots (left panels) of $x_i$ of all the dynamical units (N=200) and their spatiotemporal patterns (right panels) for different $k$ values.
A smooth profile of a coherent state is observed  in Fig.~\ref{bothsym}(a) for the array without any self-feedback ($k=0$). Figure~\ref{bothsym}(b) shows  a splitting into a coherent subpopulation in the middle and two incoherent subpopulations on both sides in symmetric locations for $k=2.0$ that we call as  chimera-like states.  
 A new subgroup of oscillators emerges in a  coherent state different from the previous coherent state and another subgroup of incoherent oscillators forms escaping from the previous coherent state. Notice that the size of the coherent subpopulation is smaller than the actual size of the perturbed subpopulation and it is an emergent property of the feedback mechanism as mentioned above.
However, the size of the coherent subpopulation in the chimera-like states decreases with $k$ (not shown here) and  finally, the whole population becomes completely incoherent for a larger $k=3.5$ as shown in Fig.~\ref{bothsym}(c).  Figures ~\ref{bothsym}(e)-(g) show spatiotemporal patterns that  corroborate the snapshots in Figs.~\ref{bothsym}(a)-(c), respectively. For a further increase in $k$-value, the network transits to another coherent state,  a partial homogeneous steady state (HSS). In this coherent state, the whole population except two  groups, each  of three oscillators $(i=1, 2, 3~~ \mbox{and} ~~ i=198, 199, 200)$ at both  ends  are  in HSS as  clear from the snapshot in Figure~\ref{bothsym}(d) for $k=4.5$ and the spatiotemporal pattern in Fig.~\ref{bothsym}(h). 
\begin{figure}[ht]
	\centering
	\includegraphics[width=0.5\textwidth]{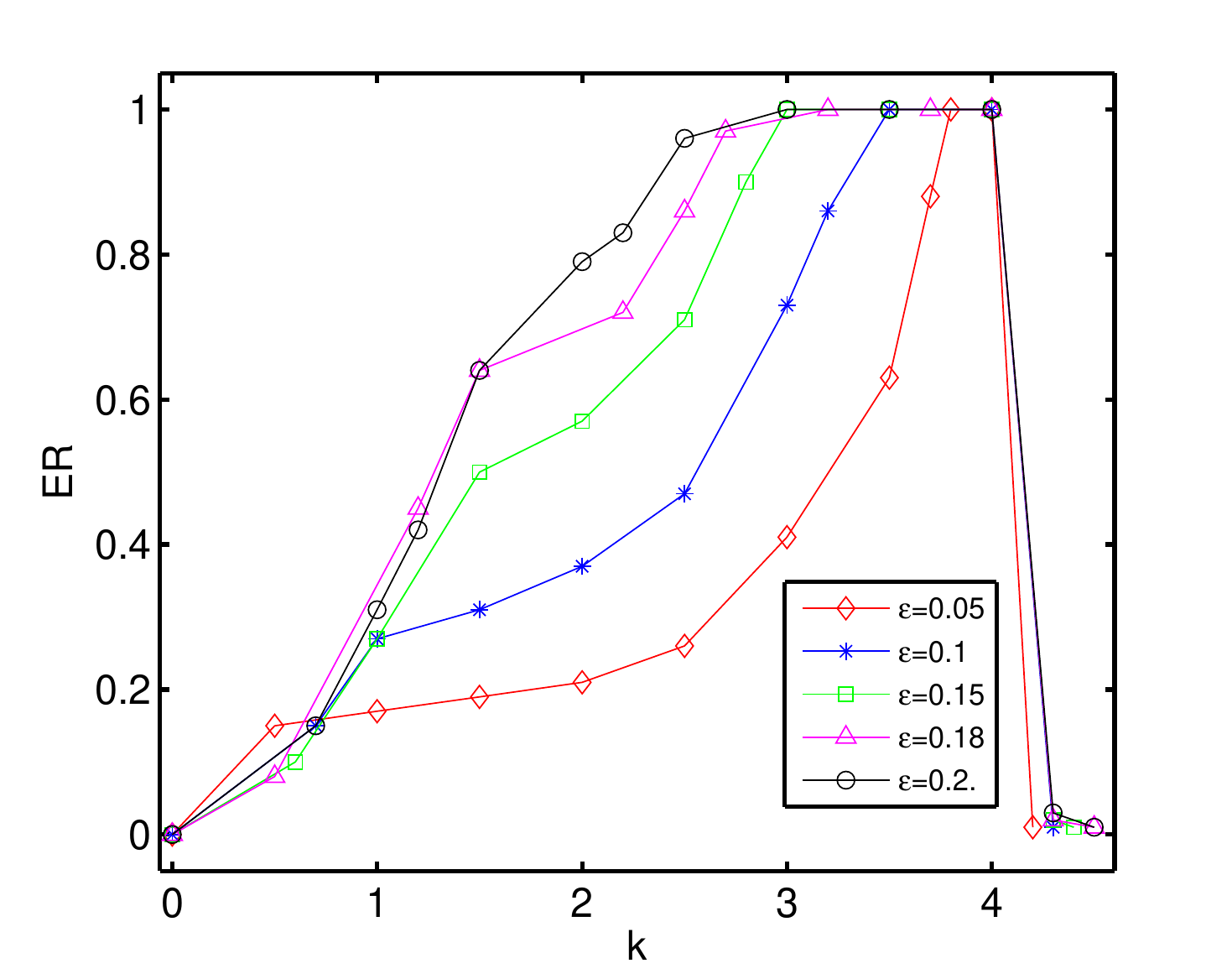} \\[-0.4cm]
	\caption{LS system: variation of escape ratio by varying the feedback strength $k$ for different local coupling strengths $\epsilon$. Here  $f(1)=f(N)=0$ and $f(i)=1$ for $i=2,3,...,N-1$. }
	\label{escape}
\end{figure}
\par A chimera-like pattern thus emerges in the chain  by the application of self-feedback when we are able to manipulate the size of the coherent and the incoherent subpopulations. A number of oscillators escapes from the coherent to the incoherent group with increasing $k$ until the whole population becomes incoherent. We calculate an {\it escape ratio} (ER) which is defined as the ratio of the number of oscillators escaped to join the incoherent subpopulation and the total number of oscillators in the array. The ER parameter is bounded in the interval [0, 1]  where ER=0 and 1 represent a coherent state and an incoherent state, respectively; $0<ER<1$ represents a coexisting state. 
 Figure~\ref{escape} shows  ER plots with  $k$  values for different $\epsilon$ values. The size of the incoherent subpopulation in the chimera-like states clearly increases with $k$ for each $\epsilon$ until the whole population collapses to a coherent state for larger $k$.
\begin{figure}[ht]
	\centering
	\includegraphics[width=0.5\textwidth]{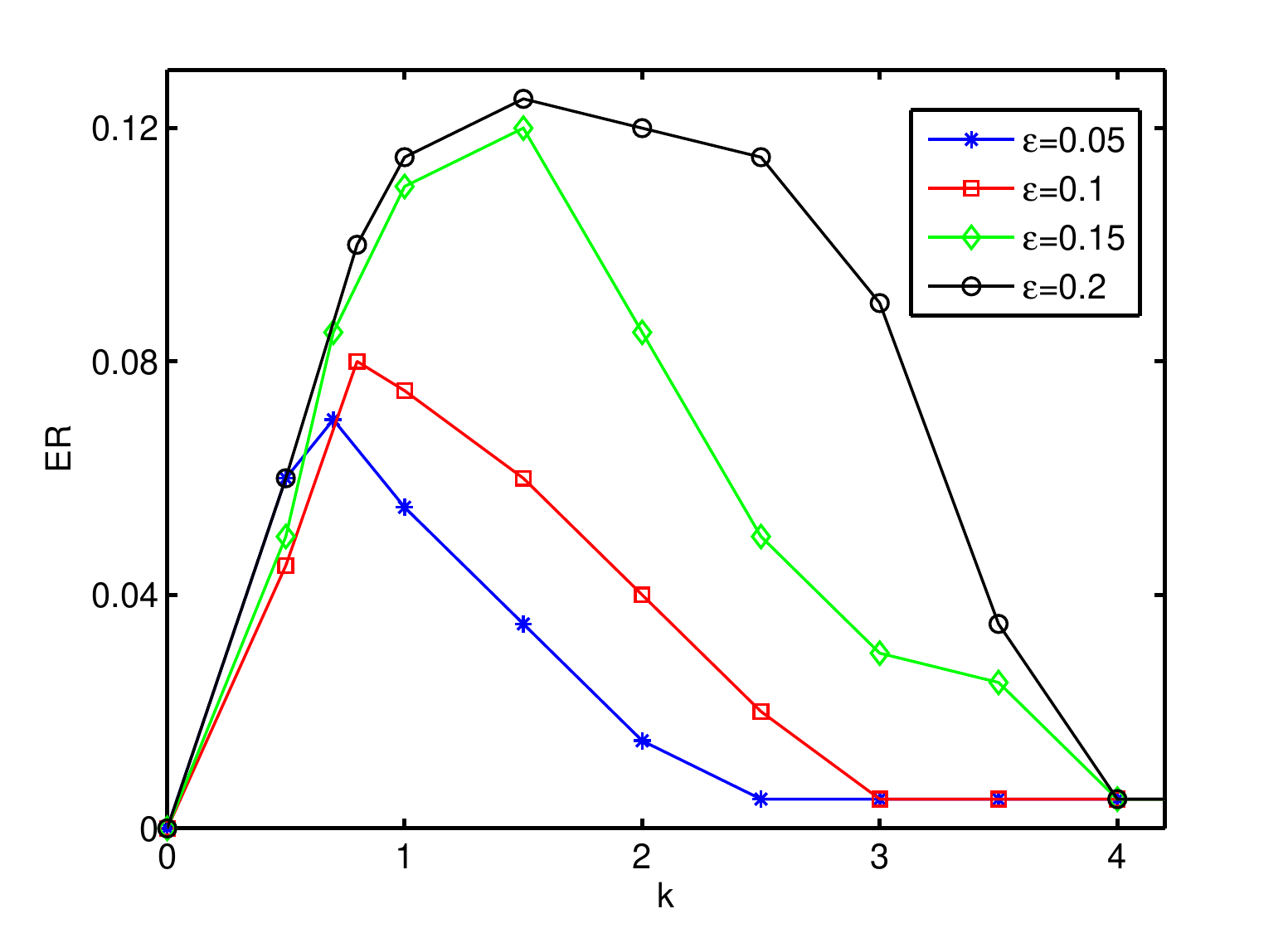} \\[-0.4cm]
	\caption{ LS system: escape ratio (ER) is plotted against the feedback strength $k$ for different values local coupling strengths $\epsilon$. Here $f(1)=1$ and $f(i)=0$ for $i=2,3,...,N$. }
	\label{escape_2}
\end{figure}
Interesting to note that larger the coupling interaction $\epsilon$ smaller is the self-feedback strength $k$ that is necessary for a complete escape of all the oscillators to the incoherent state. The ER rate with $k$ for five increasing $\epsilon$ values (red, blue, green, magenta and black colors) becomes faster  to reach at ER=1 (incoherent state). Finally, all the oscillators collapse to a coherent state for almost the same feedback strength ($k$) for different $\epsilon$ values.  At this point all the oscillators converge to HSS through saddle node bifurcation (SNB) of the coupled system except  the outer two subgroups of oscillators as mentioned above. The effect of feedback on an isolated LS oscillator is discussed in Appendix-I where the transition from oscillatory state to stable steady state emerges through SNB with the increasing value of feedback strength $k$. 
\par  Before discussing other two example cases, we address a relevant question here: what is the minimal number of oscillators needs to be perturbed to observe chimera-like pattern. To our surprise, we observe that even applying the self-feedback to a single oscillator in the chain suffices to produce the chimera-like states. Of course, it depends upon the coupling strength $\epsilon$ and feedback strength $k$ as shown in Fig.~\ref{escape_2}. We apply the self-feedback to the first oscillator ($i=1$) only of the open chain of coupled oscillators. The variation of ER is shown with feedback strength $k$ for different values of diffusive coupling strength $\epsilon$. The size of the incoherent subpopulation of the chimera-like states increases with $k$ until it reaches a maximum for critical values different for $\epsilon=0.05$ (blue), $\epsilon=0.1$ (red), $\epsilon=0.15$ (green) and $\epsilon=0.2$ (black) and  then gradually decreases until the first oscillator going to a steady state while the remaining $N-1$ dynamical units oscillate in synchronous motion. 
\par For  identification of the incoherent and the chimera-like states, we use a strength of incoherence (SI) measure \cite{rgopal} from the time series of the network dynamics. For this, we define a new  variable $w_{1,i}=x_{i+1}-x_{i}$ and $w_{2,i}=y_{i+1}-y_{i}$, which means if two neighboring oscillators are in  the coherent group then $w_{1,i}, w_{2,i}\rightarrow 0$. Next we divide the total number of oscillators into $M$ bins of equal length $n=N/M$ and then calculate the local standard deviation $\sigma_l(m)$, 
\begin{equation}
\begin{array}{lcl}
\sigma_l(m)=\langle\sqrt{\frac{1}{n}\sum_{j=n(m-1)+1}^{mn}[w_{l, j}-\langle w_l\rangle]^2}\rangle_t,
\end{array}
\end{equation}
where $m=1,...,M$ and $\langle w_l \rangle=\frac{1}{N}\sum_{i=1}^N w_{l, i}(t), l=1,2.$ For each $n$ the quantity $\sigma_l(m)$ is calculated and the SI is defined, 
\begin{equation}
\begin{array}{lcl}
\mbox{SI}=1-\frac{\sum_{m=1}^Ms_m}{M},\;\;\;\;\;\mbox{where}\;\;\;\;\;s_m=\Theta(\delta-\sigma_l(m))
\end{array}
\end{equation}
where $\Theta(\cdot )$ is the Heaviside step function and $\delta$ is a predefined threshold which is reasonably small. Consequently,  SI $=0$, SI$=1$ and $0<$ SI $<1$   represent coherent, incoherent and chimera-like states respectively. For separating chimera-like states from   multichimera-like states, we use a discontinuity measure (DM),
\begin{equation}
\begin{array}{lcl}
\mbox{DM}=\frac{\sum_{i=1}^M |s_{i+1}-s_i|}{2},\;\;\;\;\;\mbox{with}\;\;\;\;\;s_{M+1}=s_1
\end{array}
\end{equation}
Here DM is a natural number and is equal to $1$ for chimera state and a positive integer greater than $1$ i.e., $2\leq \mbox{DM} \leq \frac{M}{2}$  represents a multichimera state.

\begin{figure}[ht]
	\centering
	\includegraphics[width=0.5\textwidth]{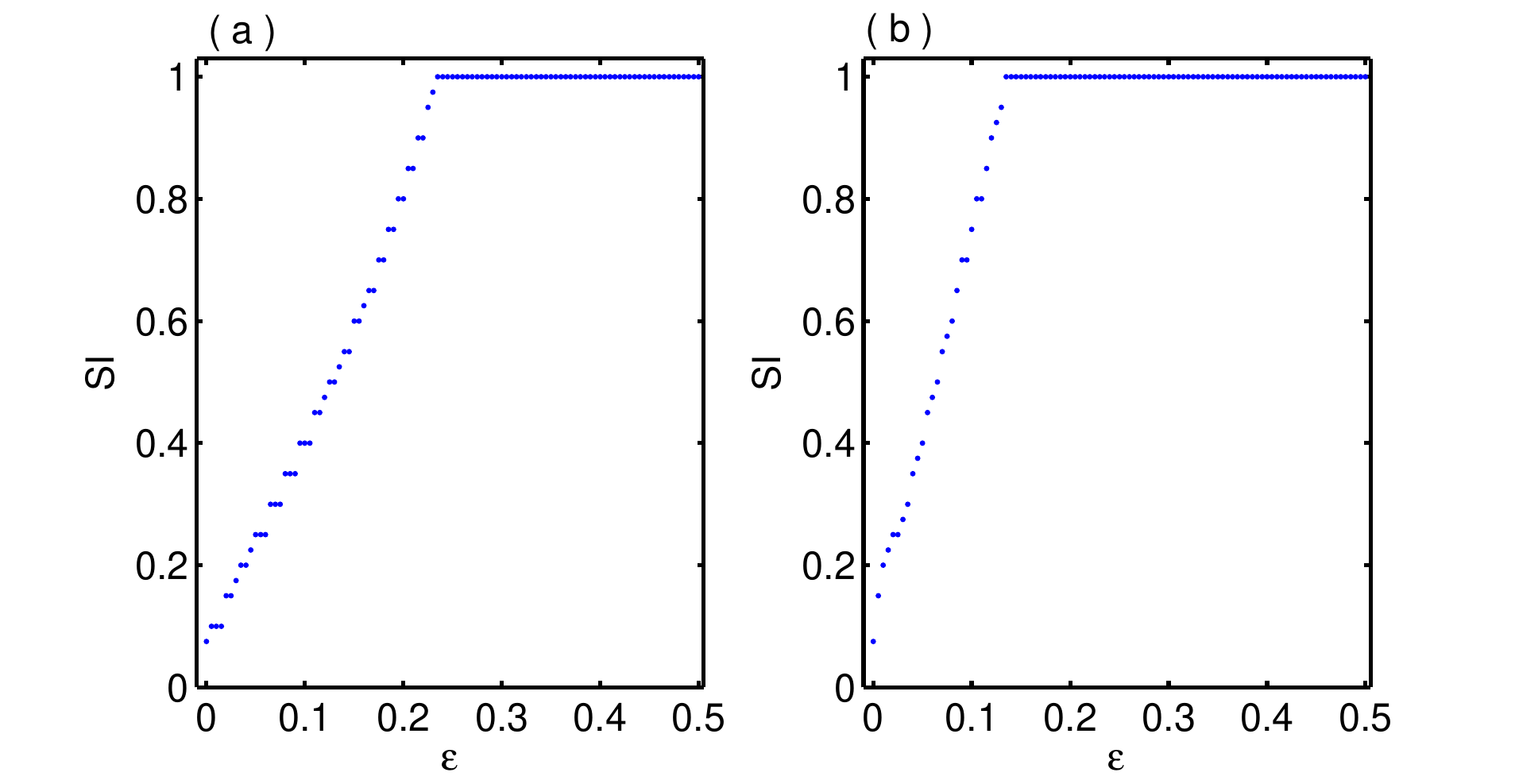} \\[-0.4cm]
	\caption{LS system: variation of SI is plotted against the coupling strength $\epsilon$ for different feedback strength (a) $k=2.0$ and (b) $k=3.0$. Here $f(1)=f(N)=0$ and all other oscillators have identical self-feedback. To calculate SI and DM, the time interval is taken over $2.5 \times 10^5$ time units after initial transient of $1 \times 10^5$ units.  Other parameters fixed at $\omega=2.0, M=20, \delta=0.05$ and $N=200$. }
	\label{SIeps}
\end{figure}

\par  We plot SI as a function of $\epsilon$ for two  feedback strengths, $k=2.0$ and  $k=3.0$, in Figs.~\ref{SIeps}(a) and \ref{SIeps}(b), respectively. A comparison with our observation in Fig.~2  reconfirms that for larger values of  the self-feedback strength $k$, the incoherent state emerges at smaller values of the  nearest neighbor coupling strength $\epsilon$. For $k=2.0$, the critical coupling strength for incoherent state is $\epsilon=0.235$ and it is smaller $\epsilon=0.135$ for larger feedback strength $k=3.0$.
\begin{figure}[ht]
	\centerline{\includegraphics[height=7cm,width=7cm]{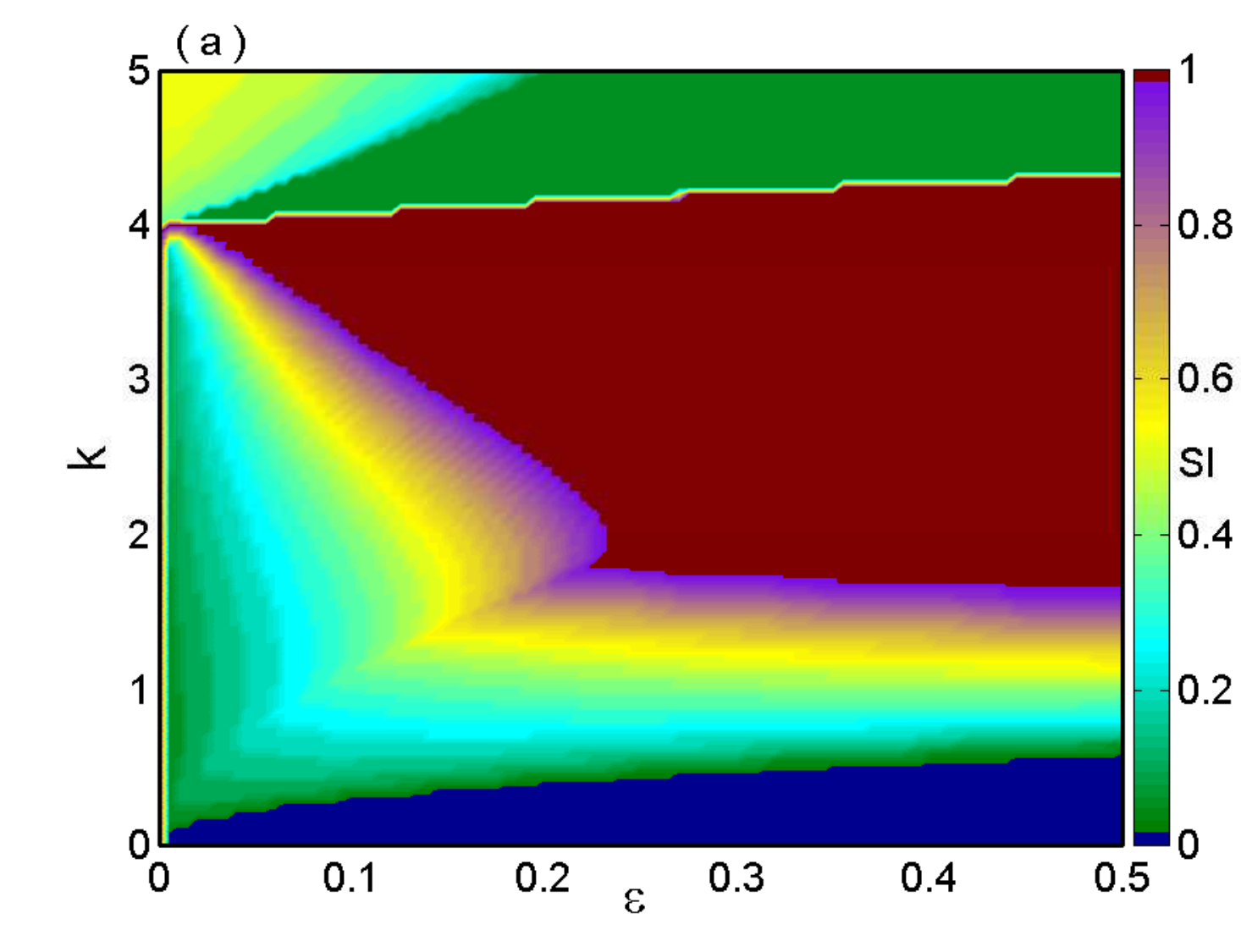}}
	\centerline{\includegraphics[height=7cm,width=7cm]{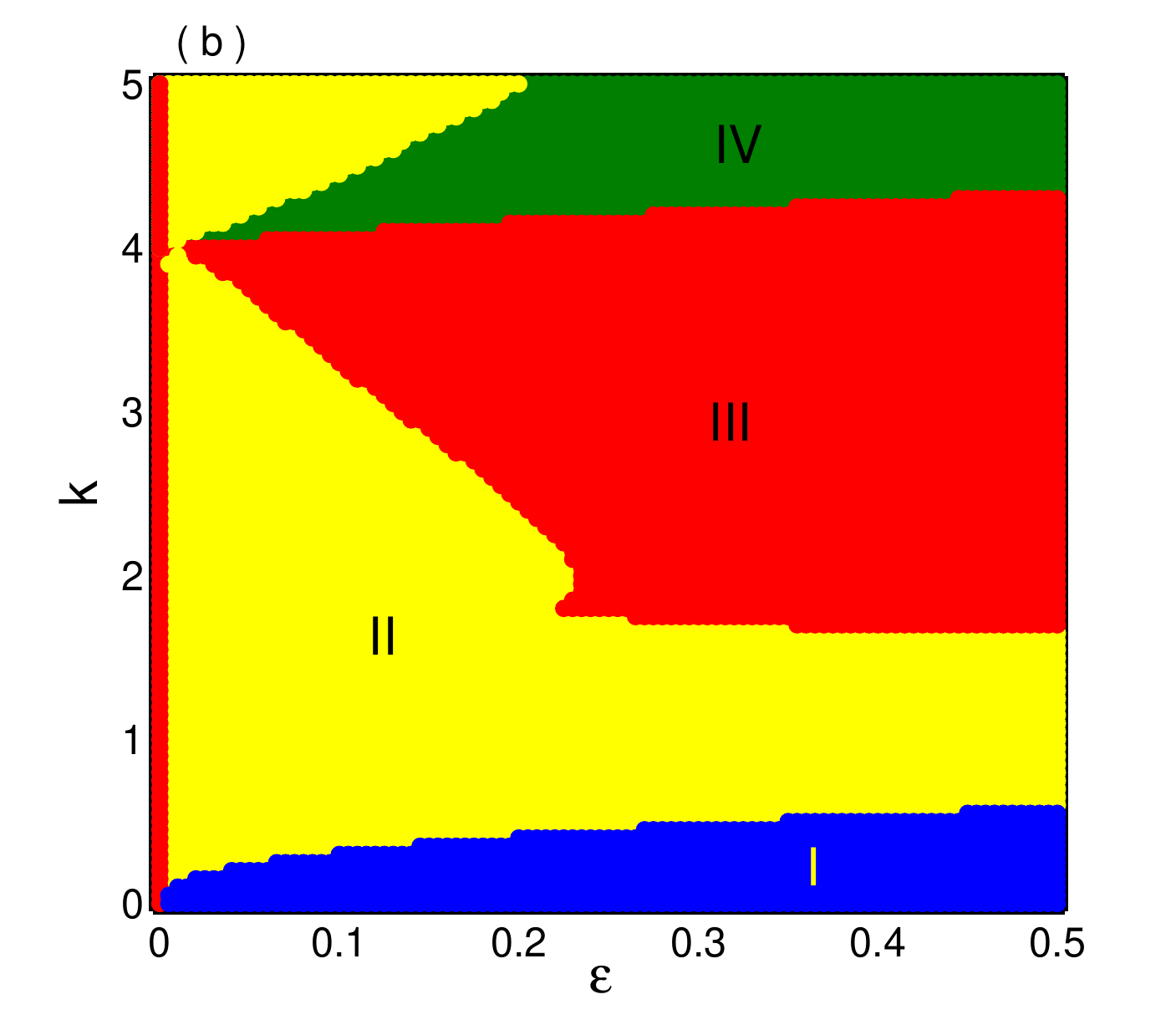}}
	\caption{LS system: two phase diagrams in $\epsilon-k$ plane showing different spatiotemporal behaviors. (a) SI is used in color bars, (b) distinguishes HSS from chimera-like states. Regions of coherent (I), chimera-like (II), incoherent (III), and homogeneous steady states (IV) are indicated by blue, yellow, red, and green color respectively. Other parameters are same as in Fig.~\ref{SIeps}.}
	\label{twopara}
\end{figure}

\par 
To capture a  broader view, at a glance, we  compute the SI values and  plot them  on a $\epsilon-k$ parameter plane in the ranges of $\epsilon \in [0, 0.5]$ and $k \in [0, 5]$ as shown in Fig.~\ref{twopara}(a). 
For incoherent and coherent dynamics, SI takes  $1$ and $0$ values which are represented by dark red and blue colors, respectively. Two regions of the chimera-like states are seen in fuzzy colors (mixing of green, cyan, yellow and violet colors).  
 Figure~\ref{twopara}(b) describes  the four distinct regions, namely, coherent (I), chimera-like states (II), incoherent (III) and HSS (IV) in the $\epsilon-k$ plane. The HSS is depicted by green color that emerges from the perturbed group of oscillators of the network and is confirmed by the spatiotemporal dynamics. 
 \begin{figure}[ht]
	\centering
	\includegraphics[width=0.5\textwidth]{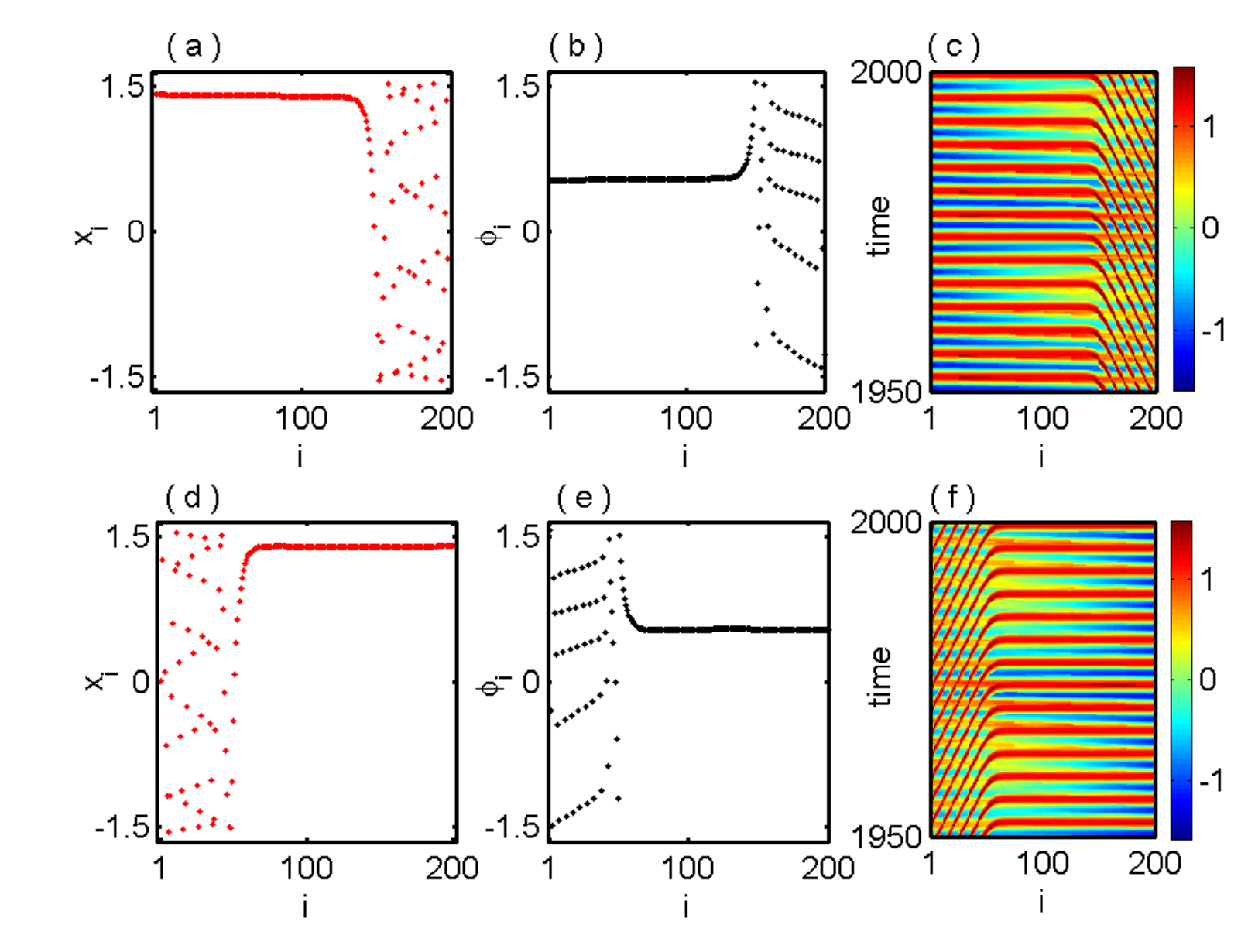} \\[-0.3cm]
	\caption{LS system: snapshot of amplitudes (left panel), corresponding snapshot of phases (middle panel) and spatiotemporal dynamics of $\phi_i$ (right panel) for fixed coupling strength $\epsilon=0.15$ and feedback strength $k=2.0$, $N=200$. The self-feedback schemes are given for situation (I)  $f(1)=f(2)=...=f(N-1)=1.0, f(N)=0$ (first row) and, situation (II) $f(1)=0, f(2)=f(3)=...=f(N)=1.0$ (second row).}
	\label{oneside}
\end{figure}

\subsection{Asymmetric feedback:  One-side population}
We consider the second case where the self-feedback is applied to a subpopulation   in one side only instead of in the middle. Two situations arise: the   self-feedback is applied identically to all the oscillators, (I) except the last one, $f(i)=1.0$ for $i=1,2,...,199$ and $f(200)=0.0$, (II)  except the first one,  $f(1)=0.0$ and $f(i)=1.0$ for $i=2,3,...,200$. In  situation (I), chimera-like states emerge in the group of identically perturbed oscillators for $k=2.0$ and $\epsilon=0.15$.  Figure~\ref{oneside}(a) shows the snapshot of $x_i$. We calculate the instantaneous phases of the oscillators using $\phi_i=tan^{-1}(\frac{y_i}{x_i})$ $(i=1, 2, ..., 200)$ and plot a snapshot of phases $\phi_i$ in Fig.~\ref{oneside}(b) where one can clearly distinguish the coherent and incoherent subpopulations in the chimera-like states. The spatiotemporal pattern of the phase variable $\phi_i$ confirms an emergence of coexisting coherent and incoherent patterns in Fig.~\ref{oneside}(c). In  the Case (II), the self-feedback is just reversed where it is applied to all the oscillators except the first one. Snapshots of $x_i$ and  $\phi_i$ and, corresponding  spatiotemporal plots of $\phi_i$ are shown in Figs.~\ref{oneside}(d)-(f) respectively. Clearly, a change of side of the perturbed node does not affect the nature of the chimera-like states or the size of the coherent and incoherent subpopulations which switches  sides. Note that the size of the incoherent or the coherent subpopulation in the chimera-like states  is smaller than the number of perturbed oscillators that confirms the emergent property of the chimera-like states. The size of two subpopulations can again be controlled  by varying $k$ as previously done.

\begin{figure}[ht]
	\centering
	\includegraphics[width=0.45\textwidth]{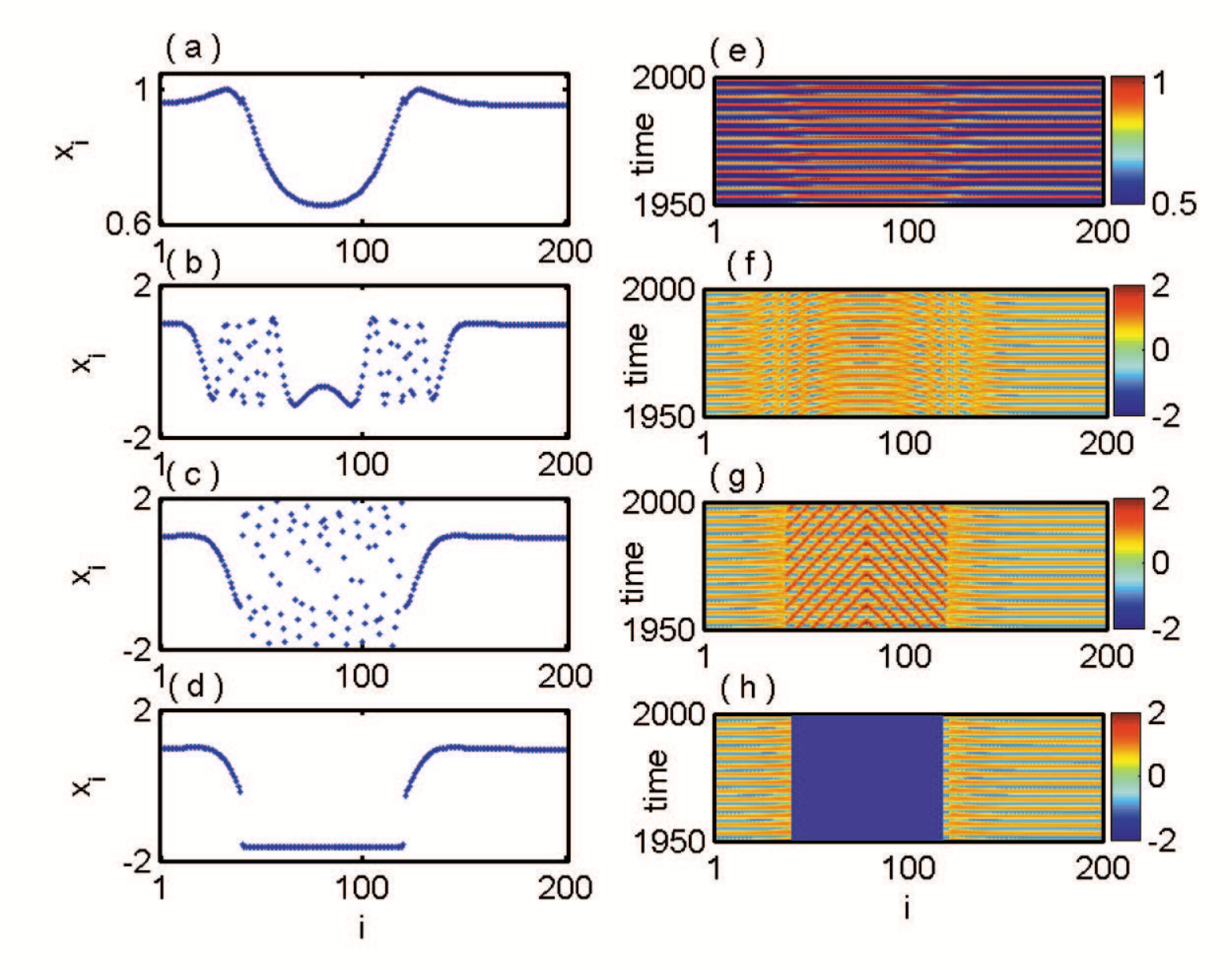} \\[-0.4cm]
	\caption{LS System: Snapshots of amplitude (left panel) and corresponding spatiotemporal dynamics (right panel): (a) coherent state, $k=0.10$, (b) chimera state, $k=0.50,$ (c) incoherent state, $k=3.5$ and (d) stable steady state, $k=4.3$. The self-feedback scheme is $f(i)=0$ for $i=1,...,40,121,...,200$ and $f(i)=1$ for $i=41,...,120$. The value of coupling strength $\epsilon=0.18$. }
	\label{asymmetric}
\end{figure}

\subsection{Asymmetric feedback: Intermediate population}
The self-feedback is applied to a group of oscillators in the middle, however, this time the number of unperturbed oscillators are not equal at the two ends. An asymmetry is created in the size of the unperturbed subpopulations. There are many possible choices of asymmetric positions of the oscillators for the application of the feedback scheme. We take one of the  choices as  $f(i)=0$ for $i=1,2,...,40$ and for $i=121,122,...,200$ while $f(i)=1$ for $i=41,42,...,120$. We fix $\epsilon=0.18$ and vary $k$; at a lower value of $k=0.1$, all the oscillators are in a coherent state in Fig.~\ref{asymmetric}(a) where a smooth asymmetric profile emerges. At a higher value of $k=0.5$, a typical multichimera-like pattern emerges where two subgroups of incoherent oscillators rest between three subgroups of coherent oscillators in Fig.~\ref{asymmetric}(b).  However, we call these as chimera-like states instead of multichimera states since the chimera-like states are restricted to the perturbed group of oscillators whereas unperturbed groups $(x_1, ...,x_{40}; x_{121},...,x_{200})$ are still in coherent state.  With a further increase of $k=3.5$, the typical chimera-like pattern emerges with one incoherent subgroup resting between two  asymmetric subgroups of coherent oscillators in Fig.~\ref{asymmetric}(c).  The original sizes of the perturbed and the unperturbed oscillators remain unchanged where the perturbed oscillators only switch to the incoherent state as expected. We do not call such states as emergent chimera states. At a higher value of $k=4.3,$ all the perturbed oscillators are found to be coherent (Fig.~\ref{asymmetric}(d)) which is different from the coherent state in Fig.~\ref{asymmetric}(a). All the perturbed oscillators, $i=41, 42, ... , 120$, emerge into a new coherent state that  represents a HSS. The unperturbed oscillators ($i=1,...,40,121,...,200$) remain in a coherent state and periodically oscillating. The  spatiotemporal dynamics are shown in Figs.~\ref{asymmetric}(e)-(h) that correspond to the snapshots in Figs.~\ref{asymmetric}(a)-(d), respectively, and confirm our  observations.
\par
\begin{figure}[ht]
	\centerline{
		\includegraphics[scale=0.5]{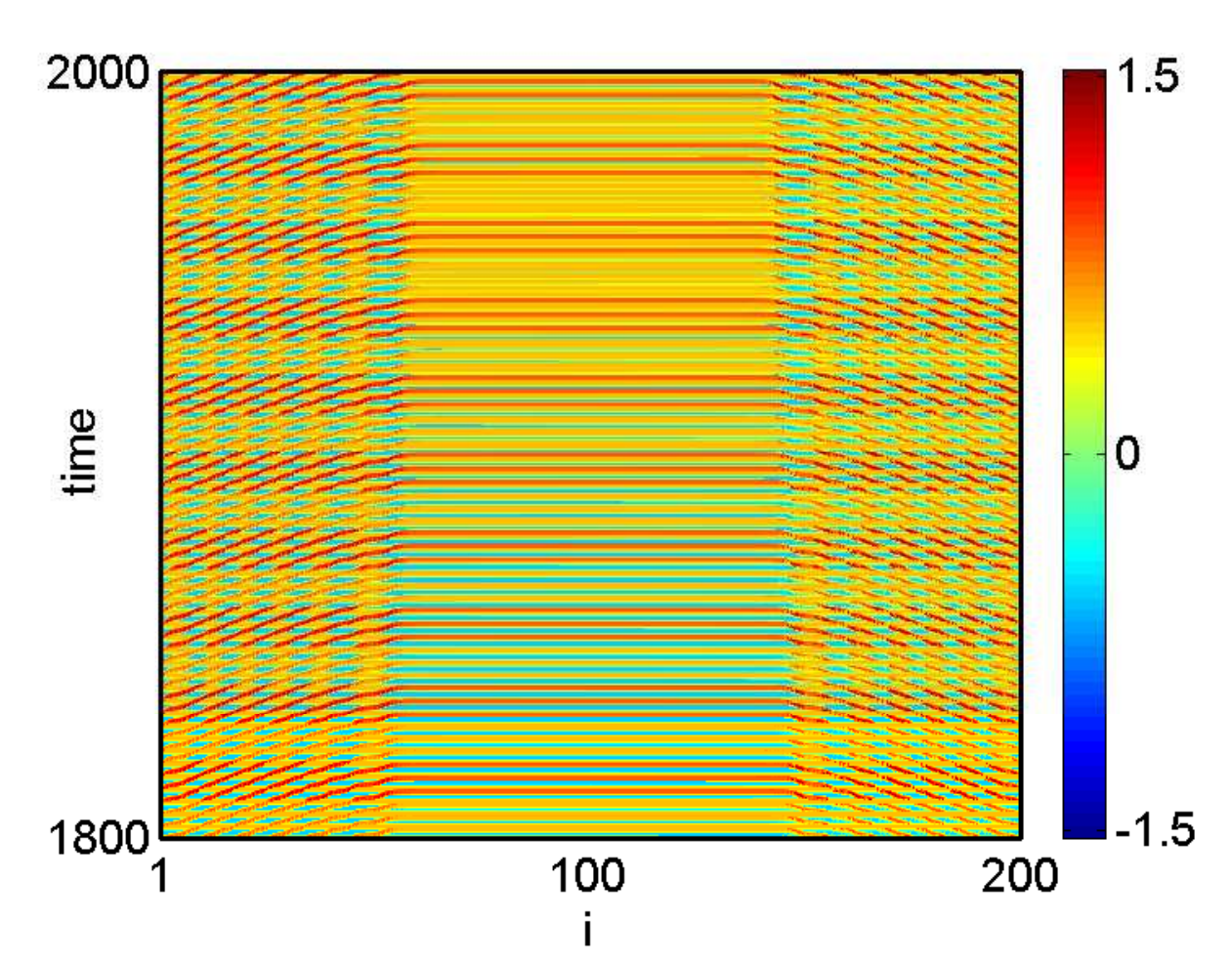}}
	\caption{ LS system: Spatiotemporal dynamics under self-feedback in regular time interval. $\omega=2.0$, $k=1.8$,  $\epsilon=0.15$, $N=200$ and $f(1)=f(N)=0$, $f(i)=1$ for $i=2,...,N-1$. }
	\label{Fig13}
\end{figure}

  Next we check the stability of the observed chimera-like state when the feedback is applied at a regular time interval instead of all time perturbation.  The self-feedback is applied to all the oscillators except first and last node of the network of locally coupled oscillator i.e. $f(1)=f(N)=0$ and $f(i)=1$ for $i=2,...,N-1$. The feedback is applied at every 1000 time units and continue up to $2 \times 10^5$ units. Figure~\ref{Fig13} represents the spatio-temporal dynamics of the open chain of locally coupled oscillators. It is observed that the chimera-like states persist over a long time for the application of self-feedback with a regular time interval. 
\begin{figure}[ht]
	\centering
	\includegraphics[width=0.5\textwidth]{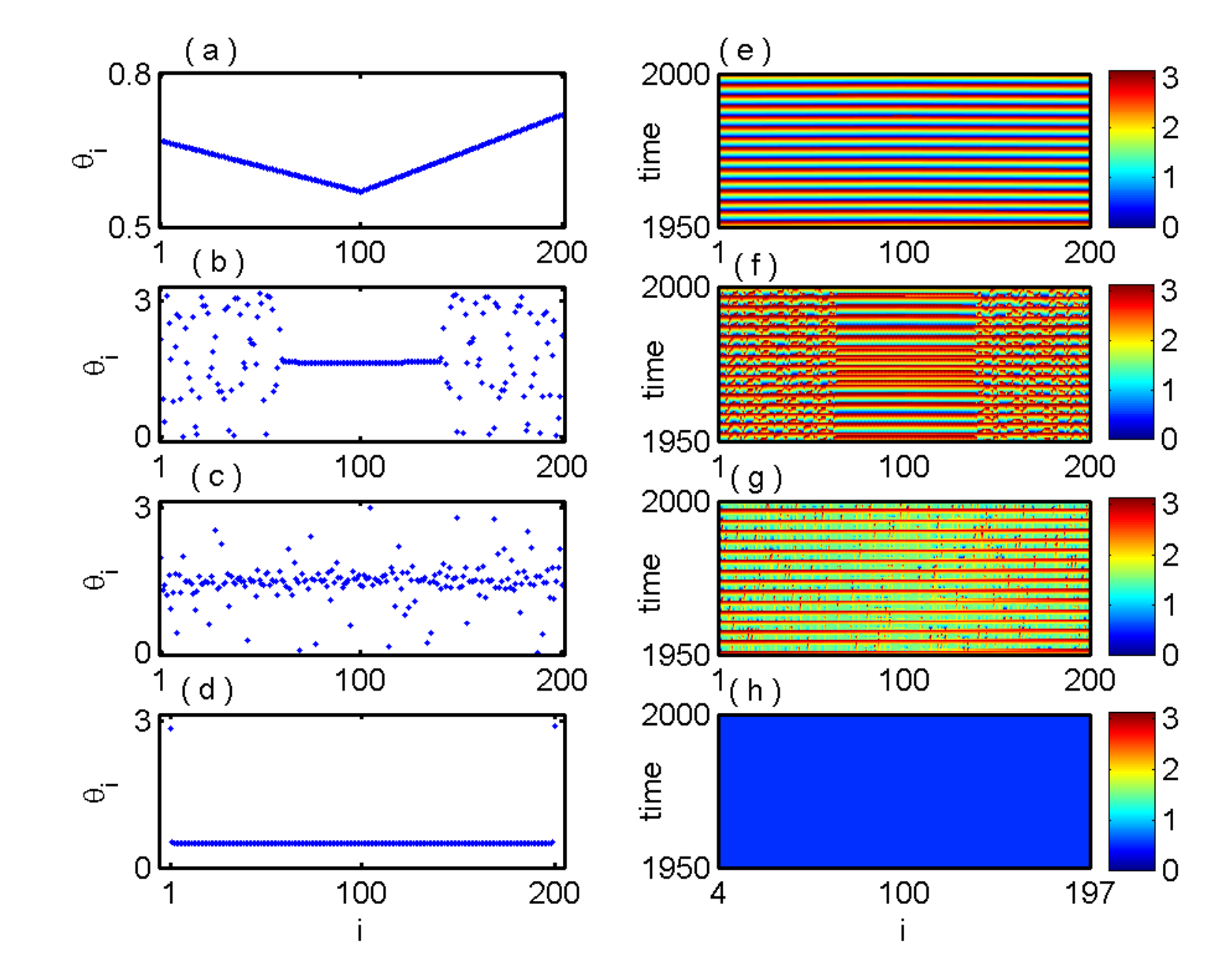} \\[-0.4cm]
	\caption{KS model: Snapshots of the phases $\phi_i(t)$ shows (a) coherent state, $k=0.0$, (b) chimera-like state, $k=0.2$, (c) incoherent state, $k=0.95$, (d) homogeneous steady state, $k=2.0$. (e-h) show the corresponding space time plot. Here $\epsilon=0.05, \alpha=1.5$ and $N=200$. }
	\label{ksfirst}
\end{figure}
\section{Network of phase oscillators}
We consider the KS phase oscillator, as a second paradigmatic model, \cite{kuramoto} to construct the  open chain network to investigate the generic property of our proposed self-feedback scheme in the emergence of  the chimera-like states. The chain of oscillators with self-feedback is
\begin{equation}
\begin{array}{lcl}
\dot \theta_1=\omega+\epsilon \sin (\theta_2-\theta_1-\alpha)+f(1)k\sin(\theta_1)\\\\
\dot \theta_i=\omega+\frac{\epsilon}{2}\{ \sin (\theta_{i+1}-\theta_i-\alpha)+\sin (\theta_{i-1}-\theta_i-\alpha)\}\\\\
\;\;\;\;\;\;\;\;\;\;\; +f(i)k\sin(\theta_i)\;\;\;\;\;\;   \mbox{for}\;\;\; i=2,...,N-1\\\\
\dot \theta_N=\omega+\epsilon \sin (\theta_{N-1}-\theta_N-\alpha)+f(N)k\sin(\theta_N).
\end{array}
\end{equation}
where $\omega =1.0$ is the  frequency, $\epsilon$ is the local coupling strength, $k$ is the self-feedback strength and $\alpha$ is a phase-lag parameter. The phase-lag plays \cite{kuramoto,strogaz_prl} a crucial role in the emergence of chimera states  in nonlocally coupled oscillators  in a range of  $\alpha\in[1.45;1.57]$.
\begin{figure}[ht]
	\centering
	\includegraphics[width=0.45\textwidth]{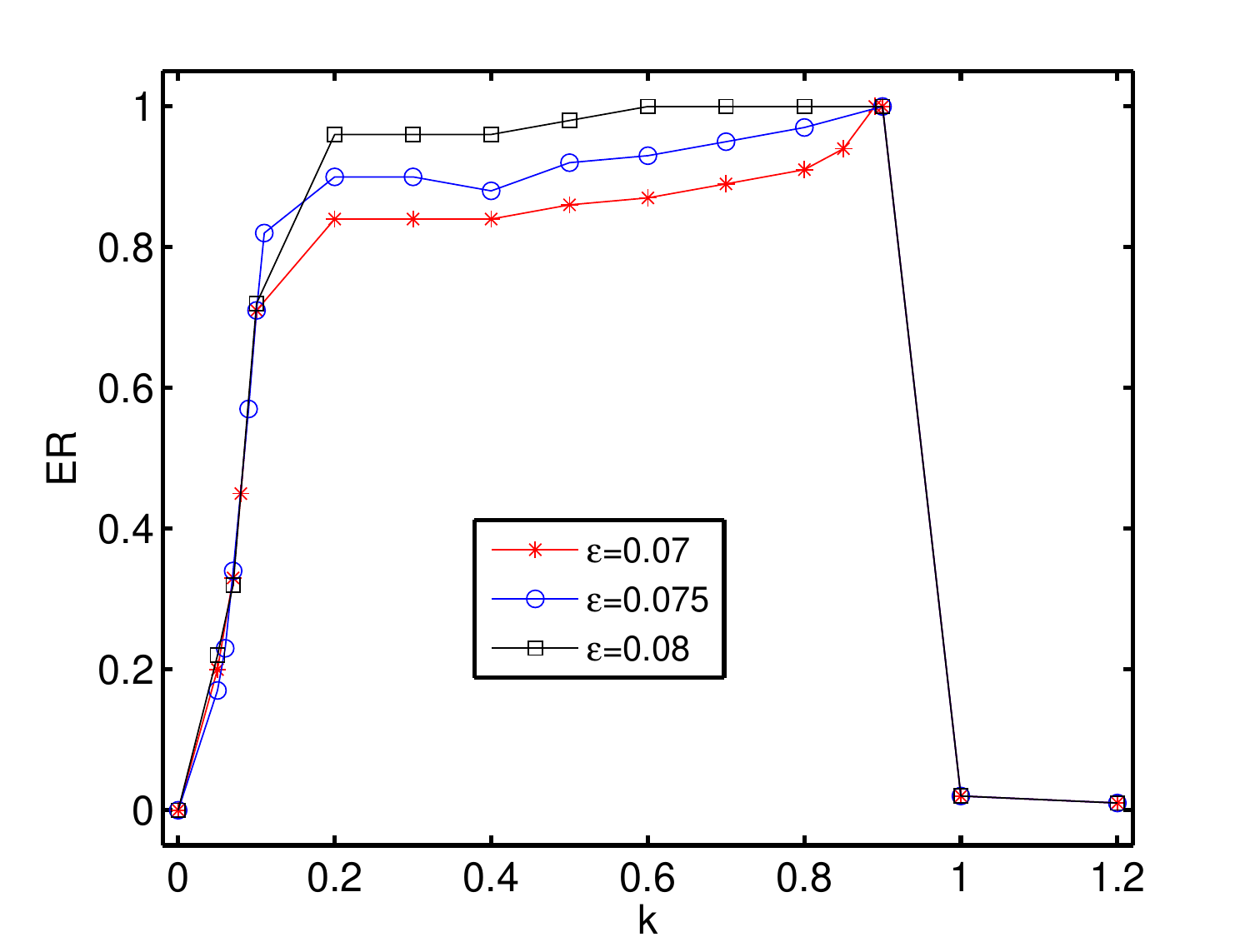} \\[-0.4cm]
	\caption{KS model: variation of escape ratio by varying the self-feedback strength $k$ for fix values of local coupling strengths $\epsilon.$ }
	\label{escapeks}
\end{figure}
\par Considering our proposed self-feedback scheme, 
we first apply the self-feedback to all the oscillators except the first and the last one as described in Eq.~(5): $f(1)=f(N)=0$ and $f(i)=1$ for $i=2,3,...,N-1$. We start with  all the oscillators  in a coherent state without any self-feedback and form a smooth profile for $\epsilon=0.05$. A snapshot of the phases $\phi_i(t)$ is shown in Fig.~\ref{ksfirst}(a). With the application of self-feedback of strength $k=0.2$, chimera-like states emerge with coexisting subpopulations of coherent and incoherent oscillators as shown in Fig.~\ref{ksfirst}(b). Similar to the case of the LS chain, a small sub-subpopulation of the perturbed subpopulation moves to a new coherent state while all the other oscillators at the two ends become incoherent. This confirms the emergent property of the chimera-like state  from the influence of partial self-feedback. With further increase of $k$, the number of coherent oscillators gradually reduces  in the chimera-like pattern and finally becomes completely incoherent for $k=0.95$ as shown in Fig.~\ref{ksfirst}(c). At a higher value of $k=2.0$, the coherent HSS appears in the perturbed subpopulation of oscillators ( $\theta_2, \theta_3,...,\theta_{199}$) as shown in Fig.~\ref{ksfirst}(d). Note that two unperturbed  oscillators at the two ends remain in coherent state. The spatiotemporal patterns of the coherent state without feedback, chimera-like states, incoherent state  and coherent state with feedback are shown in Figs.~\ref{ksfirst} (e)-(h), respectively. 
We plot the ER with changing $k$ for different $\epsilon$ as shown in Fig.~\ref{escapeks}. For all $\epsilon$ values, the
size of the incoherent population increases continuously; however, they collapse sharply at an end point where the coherent state re-emerges as usually seen in the case of the LS chain. 

\begin{figure}[ht]
	\centering
	\includegraphics[width=0.5\textwidth]{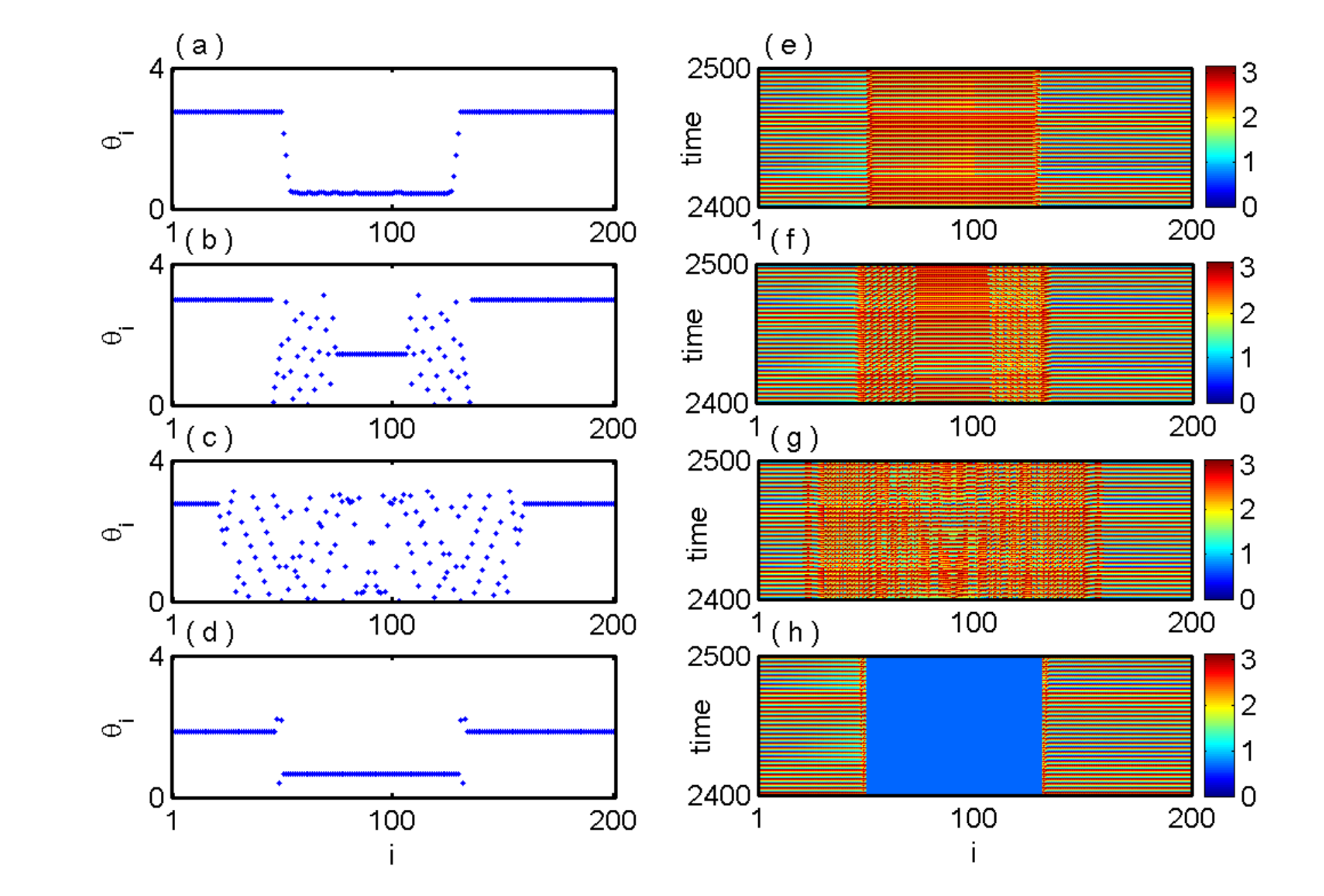} \\[-0.4cm]
	\caption{  KS model: snapshots in left panels (a)-(d)  for  varying $k=0.005, 0.1, 0.132,$ and $1.6$ and fixed $\epsilon=0.028$. Corresponding spatiotemporal behaviors  are shown in ((e), (f), (g) and (h), respectively.}
	\label{kssecond}
\end{figure}
In a second example of the KS chain, we apply identical self-feedback to a selected number of intermediate oscillators in the array. The feedback scheme is  $f(i)=0$ for $i=1,...,50,131,...,200$ and $f(i)=1$ for $i=51,...,130$ where the perturbed subpopulation is in an asymmetric location. 
One again we fix $\epsilon=0.028$ and vary $k$ when the perturbed and the unperturbed subpopulations follow two different coherent profiles as shown in Fig.~\ref{kssecond}(a) for a lower $k=0.005$. For  a larger $k=0.1$, the chimera-like pattern emerges as shown in Fig.~\ref{kssecond}(b) where two incoherent  subpopulations emerge and they extend to both the perturbed and unperturbed oscillators. A  coherent profile emerges in an intermediate subpopulation. With further increase of $k$, the number of incoherent oscillators increases and finally a large subpopulation switches to an incoherent state (Fig.~\ref{kssecond}(c)) for $k=0.132$.  For a critical $k=1.6$, the perturbed oscillators move to the HSS and the unperturbed oscillators remain coherent and periodically oscillating as shown in  (Fig.~\ref{kssecond}(d)). Figures~\ref{kssecond}(e)-(h) show the spatiotemporal patterns corresponding to Figs.~\ref{kssecond}(a)-(d), respectively.
\par
\begin{figure}[ht]
	\centering
	\includegraphics[width=0.5\textwidth]{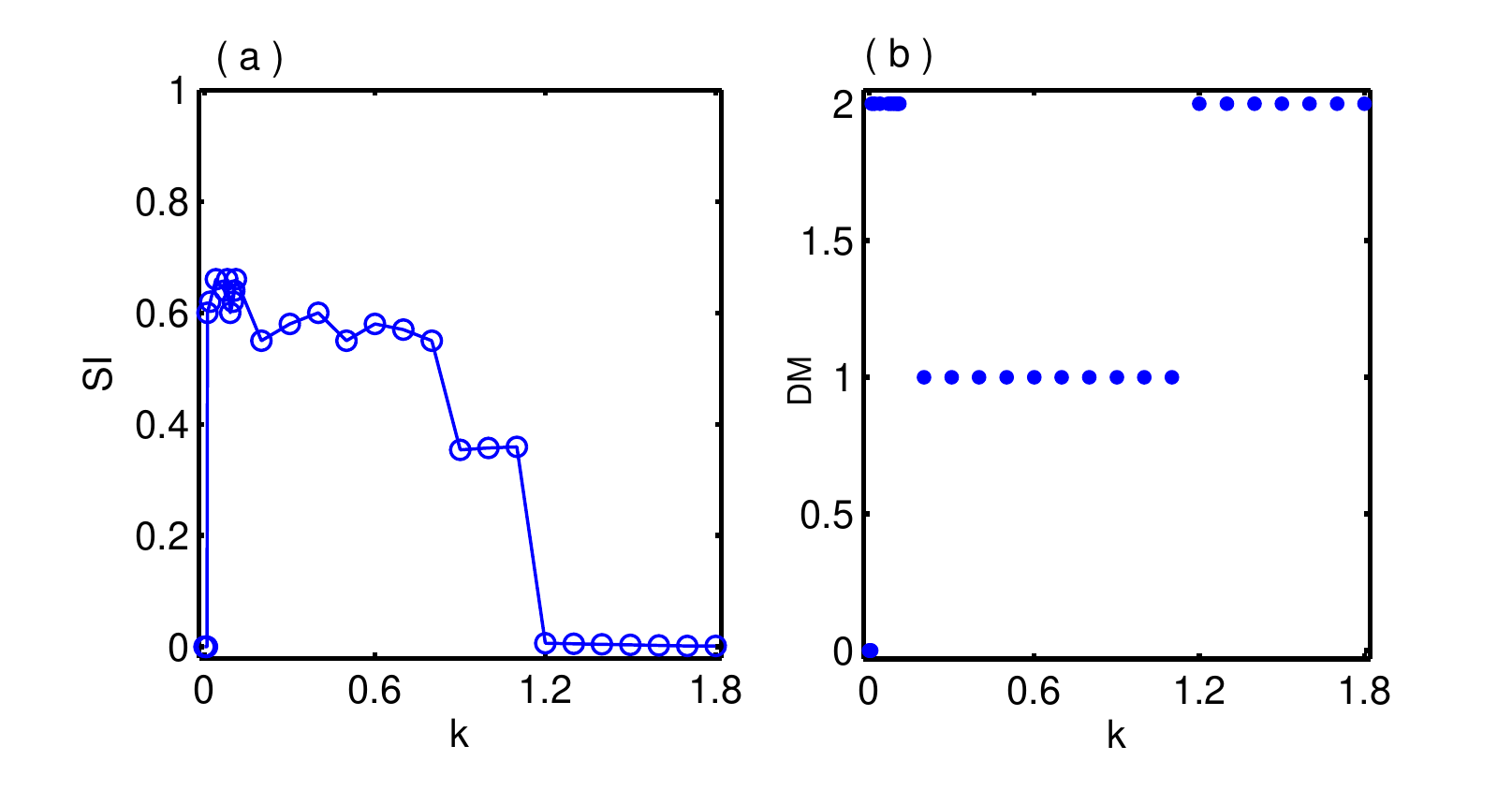} \\[-0.4cm]
	\caption{ KS model: variation of (a) strength of incoherence and (b) discontinuity measure against the feedback strength $k$ for fixed $\epsilon=0.028$.}
	\label{kssidm}
\end{figure}

 To confirm  the existence of different states in the KS chain, once again, we use the SI and the DM measures for varying $k$ and fixed $\epsilon$. As mentioned above, for the coherent state (SI, DM)=(0, 0) and (SI, DM) = (1, 0) represents an incoherent state. Furthermore, $0 <\mbox{SI}<1$, with DM = 1 and  $2\leq \mbox{DM} \leq \frac{M}{2}$ represent chimera and multichimera states, respectively.  Figures~\ref{kssidm}(a) and \ref{kssidm}(b) show the variation of SI and DM values for a fixed value of $\epsilon=0.028$ and varying  $k$. In  the absence of any feedback strength ($k=0$), all the oscillators follow a coherent profile for fixed $\epsilon=0.028$ and it persists for a small range of weak self-feedback up to  $k\simeq0.01$. Then multichimera-like state is observed in the whole network in the region of $\{k : 0.01<k\leq 0.11\}$ as SI$\in(0,1)$ and DM$=2$. For a  slightly larger value of $k$, chimera-like states occur in the region $\{k : 0.11<k\leq 1.1\}$ as SI$\in(0,1)$ and DM$=1$. At a higher value of $k$, the multichimera-like pattern is observed  again in the region of $\{k : 1.1<k\leq 1.82\}$.

\section{Conclusion}
We  explored collective dynamics in one-dimensional open chain of locally coupled oscillators  under a self-feedback scheme. We numerically observed emergent chimera-like patterns in the chain using the LS system and the KS phase model as dynamical nodes and by applying self-feedback to a selected subpopulation at different locations of a chain. The spatiotemporal  pattern was observed in a wide range of parameter space of local  coupling strength and self-feedback strength  as illustrated by phase diagrams. We identified that the chimera-like states emerge via the escape of  a number of oscillators  from the perturbed set to incoherent state. This transition was initiated by an escape of one or few oscillators from  the  coherent state with changing  self-feedback strength. With the increase of self-feedback strength, more oscillators escape from the perturbed subpopulation to join the unperturbed subpopulation and finally become completely incoherent. The size of the coherent and the incoherent subpopulations in the chimera-like states was thereby controllable by the self-feedback. The proposed feedback scheme is simple and can be implemented experimentally easily. It may be of practical utility for engineering and control of coexisting coherent and incoherent patterns in other homogeneous or partially homogeneous networks.

\section{acknowledgments}
B.K.B, D.G. and S.K.D. were supported by SERB-DST (Department of Science and Technology), Government of India (Project no. INT/RUS/RFBR/P-181). G.V.O. was supported by the Russian Foundation for Basic Research (Grant 15-52-45003) and by the Russian Science Foundation (Grant 14-12-00811). S.K.D. also acknowledges support by the Emeritus Fellowship of the University Grants Commission (India).

\section{Appendix I: Single Landau-Stuart oscillator under self-feedback}
The dynamical equation of the individual LS oscillator in presence of self-feedback is given by
\begin{equation}
\begin{array}{lcl}

\dot x=(1-x^2-y^2)x-\omega y+kx\\
\dot y=(1-x^2-y^2)y+\omega x.

\end{array}
\end{equation}
 where $\omega$ and $k$ are the  frequency and the feedback strength, respectively. Besides the trivial fixed point $(0,0)$, the system has four more fixed points that are feedback strength dependent as given by, $x^*=-\frac{\omega y^*}{\omega^2-ky^{*2}}$ where $y^*=\pm \frac{1}{\sqrt{2k}}\sqrt{(k+2\omega^2)\pm \sqrt{(k+2\omega^2)^2-4(k\omega^2+\omega^2+\omega^4)}}$. The (0,0) fixed point is always unstable since the eigenvalues $\lambda_{1,2}=(1+\frac{k}{2})\pm \frac{k}{2}\sqrt{1-\frac{4\omega^2}{k^2}}$ of the Jacobian at $(0,0)$ are always positive. 

\begin{figure}[ht]
\centerline{
\includegraphics[scale=0.5]{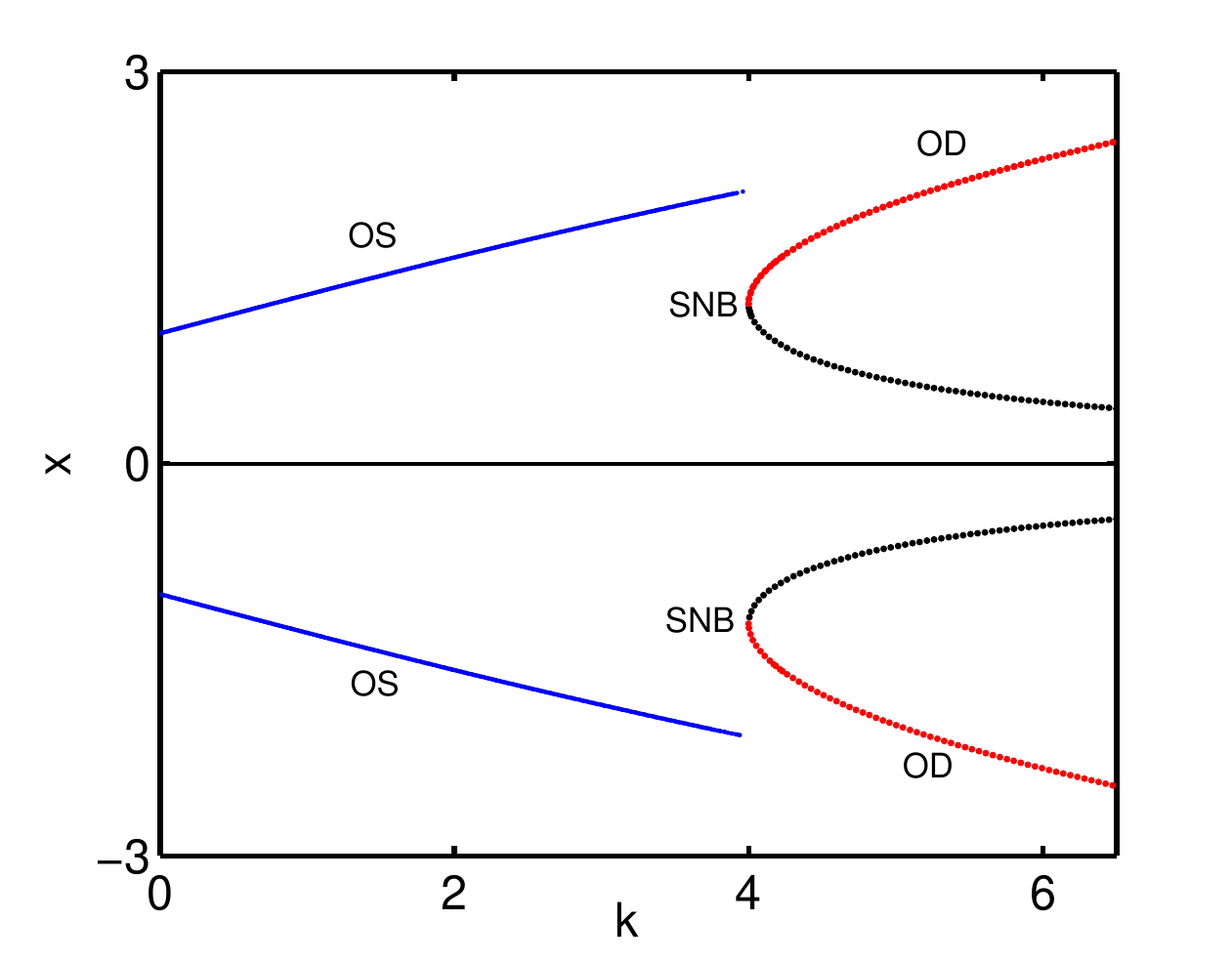}}
\caption{ LS system: maximum of $x$ is plotted against the feedback strength $k$ by fixing  $\omega=2$. Blue, black and red lines indicate the oscillatory state (OS), unstable fixed points and stable fixed points (OD), respectively. }
\label{Fig12}
\end{figure}

Figure~\ref{Fig12} shows a bifurcation diagram of an isolated LS oscillator for  varying feedback strength $k$ with a fixed $\omega=2$. For lower values of $k$, the system exhibits a limit cycle oscillation (OS) when all five  fixed points are unstable and the oscillation persists up to a threshold value $k\lesssim 4.0$. For  ($k\gtrsim$4.0), the oscillation ceases to two different stable points via  saddle node bifurcation (SNB) while the other fixed points including the trivial (0,0) fixed point still remain unstable.\\\\

\end{document}